\documentclass[runningheads]{llncs}
\usepackage{afterpage}
\usepackage{algorithmic}
\usepackage{array} 
\usepackage{amsmath,amssymb,amsfonts}
\usepackage{booktabs} 
\usepackage{color}
\usepackage{graphbox}
\usepackage{graphicx}
\usepackage{enumitem}

\usepackage{mathtools}
\usepackage{multirow}
\usepackage{tabularx}
\usepackage{textcomp}
\usepackage{tikz}
\usepackage{pifont} 
\usepackage{colortbl}
\usepackage[normalem]{ulem}
\useunder{\uline}{\ul}{}
\usepackage{hyperref}
\usepackage{cleveref} 
\usepackage{atbegshi} 

\usepackage{bibentry}
\usepackage{longtable}
\usepackage{csquotes}

\usepackage{flushend}
\usepackage{cleveref}
\usepackage{subfig}
\usepackage{graphicx}
\usepackage{orcidlink}
\usepackage{changepage}

\usepackage{pgfplots}
\usepackage{src/tkz-kiviat} 
\usetikzlibrary{arrows}

\usepackage[nomessages]{fp}
\usepackage{censor}
\usepackage[font=small,skip=6pt]{caption}
\usepackage{lmodern}
\usepackage{relsize}

\newcounter{ch}
\newcommand{\eu}{
Funded by the European Union (TEADAL, 101070186).}

\newcommand{\sul}{\rule{0.15cm}{0.15mm}}
\usepackage{array,ragged2e}
\newcolumntype{P}[1]{>{\RaggedRight\arraybackslash}p{#1}}
\usepackage[T1]{fontenc}
\usepackage{algorithm,algorithmic}

\AtBeginShipout{\ifnum\value{page}=16\pagecolor{red!20}\fi}
\begin{document}
\title{Designing Reconfigurable Intelligent Systems with Markov Blankets\thanks{
\eu{}}}
\titlerunning{Designing Reconfigurable Intelligent Systems with Markov Blankets}
%

\author{Boris Sedlak\orcidlink{0009-0001-2365-8265} \and
Victor Casamayor Pujol\orcidlink{0000-0003-2830-8368} \and
Praveen Kumar Donta\orcidlink{0000-0002-8233-6071} \and
Schahram Dustdar\orcidlink{0000-0001-6872-8821}}

\authorrunning{Sedlak et al.}
\institute{Distributed Systems Group, TU Wien, Vienna 1040, Austria.\\
\email{\{b.sedlak, v.casamayor, pdonta, dustdar\}@dsg.tuwien.ac.at}}
\maketitle              
\begin{abstract}
Compute Continuum (CC) systems comprise a vast number of devices distributed over computational tiers. Evaluating business requirements, i.e., Service Level Objectives (SLOs), requires collecting data from all those devices; if SLOs are violated, devices must be reconfigured to ensure correct operation. If done centrally, this dramatically increases the number of devices and variables that must be considered, while creating an enormous communication overhead.
To address this, we (1) introduce a causality filter based on Markov blankets (MB) that limits the number of variables that each device must track, (2) evaluate SLOs decentralized on a device basis, and (3) infer optimal device configuration for fulfilling SLOs.
We evaluated our methodology by analyzing video stream transformations and providing device configurations that ensure the Quality of Service (QoS). 
The devices thus perceived their environment and acted accordingly -- a form of decentralized intelligence.


\end{abstract}

\section{Introduction}\label{sec:intro}
Computing Continuum (CC) systems as envisioned in~\cite{Beckman2020,dustdar_distributed_2023} are large-scale distributed systems composed of a wide variety of devices. Applications running in the CC pose ambitious requirements, e.g., near real-time latency while dealing with huge volumes of data.
Additionally, requirements may change over time; to provide the best possible service, the CC system must adapt.
However, given the highly distributed nature of the CC, it is a challenging task to dynamically reconfigure all contained devices, while ensuring high-level system objectives.

In this regard, we envision CC systems employing decentralized intelligence, which allows system parts to make decisions independently, in favor of the application running on top.
Smaller units in the CC (e.g., edge devices) would thus obtain the ability to evaluate their own state to ensure requirements are fulfilled.
One promising option to model this, is the behavioral concept introduced by Friston et al.~\cite{friston_life_2013,kirchhoff_markov_2018}. Essentially, it comprises sensory information and actions within a Markov blanket (MB)~\cite{pearl_probabilistic_1988}, through which a \textit{thing} interacts with its environment.
%
%
The MB shields the \textit{thing} from all the variables it is conditionally independent of.
Therefore, to determine the state of the \textit{thing}, only the variables in the MB must be considered.
Transferring this concept to the CC, you could model each device's behavior through MBs~\cite{sedlak_controlling_2023_short} and evaluate device requirements by considering a limited amount of variables.
Existing work \cite{furst_elastic_2018,nastic_sloc_2020,sedlak_controlling_2023_short}, however, assumes prior knowledge of how metrics are related to the system state; this approach is not scalable if requirements change during operation or metric correlations are unknown at design time.
Thus, existing approaches would fail to ensure the intricate requirements of CC systems.

Each tier in the CC poses its own requirements, which must be fulfilled to create a composable and unified service. To model requirements, Cloud Computing introduced Service Level Objectives (SLOs) as a means to achieve business agreements between infrastructure provider and application developer.
However, we propose to expand SLOs to requirements that directly influence the system behavior and the application performance.
Inspired by the work of Friston et al., and continuing the research agenda set in \cite{casamayor_pujol_towards_2021,dustdar_distributed_2023}, we aim to leverage the behavioral concept of MBs to represent SLOs throughout the CC.
The causality filter of the MB reduces the scope of variables that each device must analyze; thus, decreasing the computational effort of analysis. This empowers resource-constrained devices along the Edge to evaluate SLOs themselves. 

In this paper, we propose to evaluate application requirements through MB-based SLOs. The method constrains each SLO to a set of metrics and infers configurations that fulfill them. Further, the output is explainable due to the graphical model used. 
Hence, the contributions of this article are the following:
\begin{itemize}
    \item A statistical reasoning model for analyzing conditional dependencies between metrics in distributed systems. Whenever requirements change, the model may thus itself answer which metrics are related to their fulfillment.
    \item The graphical representation of the device state as MB, which allows interpreting the device behavior. The state can be broken down into several SLOs; in case any of them is violated, it can be explained why.
    \item A mechanism to infer optimal configurations from MBs given mutable system requirements. It was evaluated under two scenarios in which our approach provided the only configuration that did not violate any SLO.
\end{itemize}

\section{Methodology}\label{sec:methodology}
\begin{figure}[!t]
    \centering
    \includegraphics[width=1.0\columnwidth]{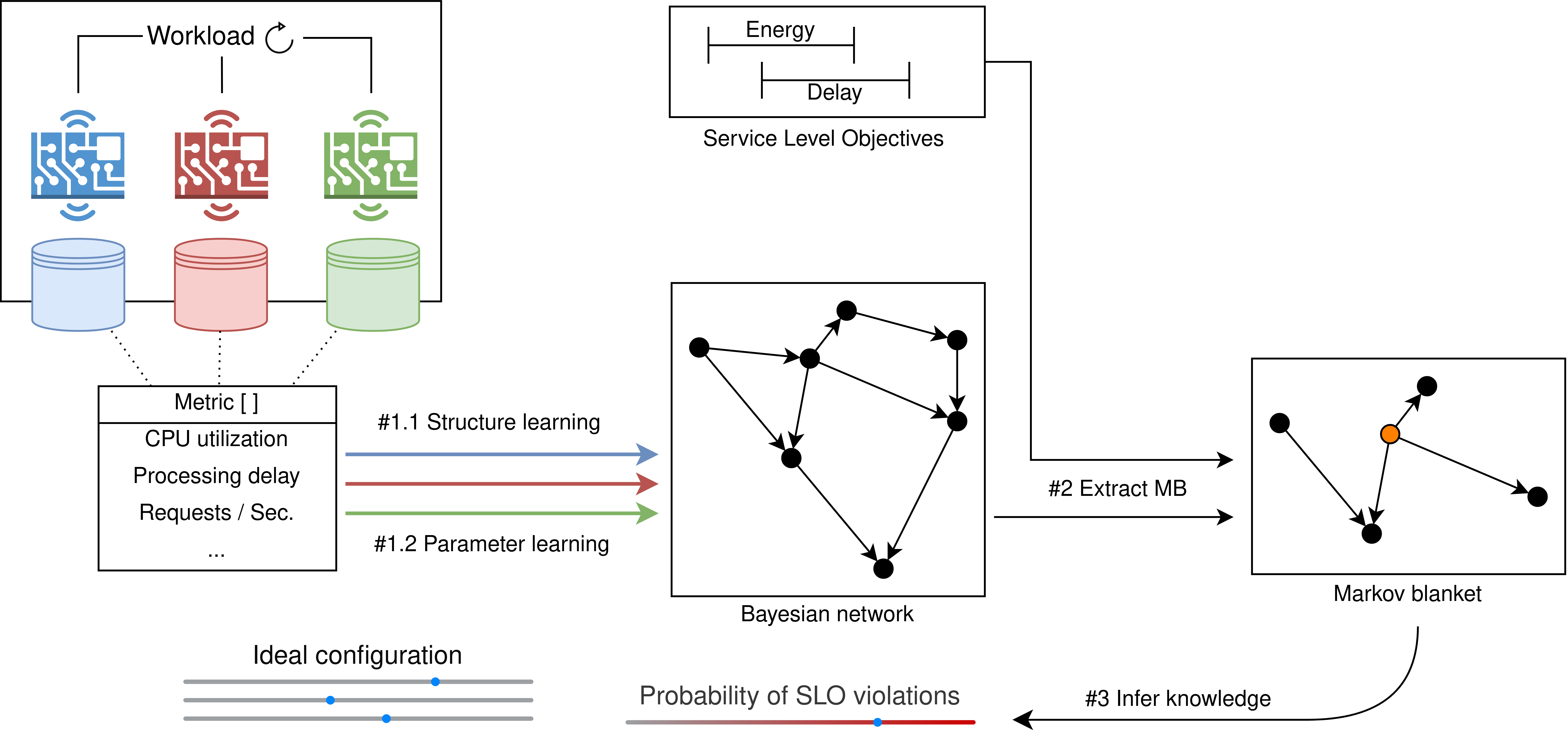}
    \caption{Methodology for training Bayesian networks and extracting knowledge}
    \label{fig:overview}
\vspace{-10pt}
\end{figure}

From a high-level perspective, we plan to analyze the device state, map selected variables to the SLO fulfillment, and provide adaptive device configurations.
Our three-step methodology to achieve this is visualized in Figure~\ref{fig:overview}:
Edge devices produce metrics about ongoing processing; then Bayesian Network Learning (\#1) is used to identify correlations between metrics and reflect the impact of environmental changes (e.g., increased incoming requests). Next, we introduce system requirements (i.e., SLOs) and extract a minimum subset of metrics for SLO fulfillment (\#2). Ultimately, we use these MBs to estimate the probability of SLO violations and (\#3) infer the configuration with the highest compliance level.

While the proposed methodology describes a sequence of actions, the tools themselves (e.g., algorithms for structure learning) can be optimized depending on the data.
This three-step methodology will be our main mechanism for predicting the probabilities of SLO violations given a device configuration.
If an SLO is violated due to an environmental change, e.g., a high request count and thus exceeded application delay, we compare possible configurations and provide the one with the highest probability of fulfilling the SLO.
This matches our envisioned level of intelligence, i.e., "understanding a situation and reacting according to needs", and neatly fits the principles of elastic computing \cite{dustdar_principles_2011}.

\section{Case Study}\label{sec:casestudy}
The following case study will be used to evaluate our methodology.
In particular, we present two video streaming scenarios that require privacy-preserving transformations. We analyze device metrics to build a Bayesian Network (BN), specify SLOs that characterize the QoS, extract the MB around each SLO, and finally, infer system configurations that have the lowest chance of violating SLOs.

\subsection{Setup}
\label{subsec:setup}

Training a BN requires data; therefore, we use the framework introduced in \cite{sedlak_privacy_2023}, which allows edge devices to detect privacy-violating patterns (e.g., screen, face, or voice) in a stream and transform it continuously to resolve possible privacy violations.
As a workload, it fits our methodology because it (1) provides an ample set of metrics reflecting the QoS of ongoing processing, (2) can be parameterized, and (3) can be executed on edge devices.
Using the framework, we specify a privacy model that detects faces within a video stream and blurs the respective region, a scenario useful for office monitoring or AR setups \cite{baniya_short_2020,sedlak_privacy_2023}.
%
During execution, 11 metrics are captured, which we introduce in Table~\ref{tab:metric-list}.
Each row contains a short description, the measurement unit, and if it can be parameterized. For example, \textit{pixel} and \textit{fps} are video stream properties; however, the producer can adapt them to create a variable \textit{bitrate}. \textit{Config} determines the device operation mode; devices such as Nvidia Jetson Xavier NX\footnotemark can thus limit their energy consumption and the number of active CPU cores. It is worth mentioning the metric \textit{distance}, which tracks the relative position of a detected face between frames, indicating how fluent/sluggish an object is tracked.

\footnotetext{https://docs.nvidia.com/jetson/archives/r34.1/DeveloperGuide/text/SO/JetsonXa\\vierNxSeries.html, accessed June 13th 2023}

\setlength{\tabcolsep}{5pt}
\begin{table}[t!]
\vspace{-4pt}
  \centering
  \caption{(Parameterizable) Metrics captured during workload execution}
  \label{tab:metric-list}
  \begin{tabular}{lclc}
    \toprule
    Name   & Unit & Description &  Param  \\
    \midrule
    \textit{delay}           & ms        & processing time per frame  & No \\
    \textit{CPU}             & \%        & utilization of the CPU  & No \\
    \textit{memory}          & \%        & utilization of the system memory  & No \\
    \textit{pixel}           & num       & number of pixel contained in a frame  & Yes \\
    \textit{fps}             & num       & number of frames received per second  & Yes \\
    \textit{bitrate}         & num       & number of pixels transferred per second  & No \\
    \textit{distance}        & px        & relative distance of object between frames  & No \\
    \textit{transformed}     & T/F       & if the model detected a pattern (i.e., face)  & No \\
    \textit{GPU}             & T/F       & if the device employs a GPU  & No \\
    \textit{config}          & nominal   & mode in which the device operates  & Yes \\
    \textit{consumption}     & W         & energy pulled by the device  & No \\
    \bottomrule
  \end{tabular}
\vspace{-4pt}
\end{table}

To explore correlations between metrics, we simulate an adaptive bitrate; precisely, the producer periodically switches between different \textit{fps} (12, 16, 20, 26, 30) and \textit{pixel} (120p, 180p, 240p, 360p, 480p, 720p), while the edge device moves through \textit{config} modes. Current parameter assignments are part of the metrics set, which is persisted with every processed frame. Metrics are directly observable by the device; except for \textit{consumption}, which is captured through an external power plug\footnotemark over a telemetry period of 10 seconds.
Metrics are accumulated in a CSV file, which will contain 756,000 rows, captured within 2.5 hours.

\footnotetext{https://www.delock.com/produkt/11827/merkmale.html, accessed June 13th 2023}

We identified five SLOs that describe the system state in terms of QoS and Quality of Experience (QoE); however, each applicable scenario can have its own subset of relevant SLOs. We assign a name to each SLO and highlight the metrics from Table~\ref{tab:metric-list} (e.g., \textit{bitrate}) that are used to evaluate the state of the SLO. 
Some SLOs are constructed by combining metrics (i.e., \textbf{within\sul time}), others are compared against a customizable threshold (e.g., \textbf{pixel\sul distance}), while other SLOs directly mimic the value (True/False) of the metric (i.e., \textbf{transform\sul success}). 


    \begin{description}
          \item[\textbf{network\sul usage}] Edge devices have limited network interfaces, and in some cases, limited network bandwidth. Since video streams are transferred over the network, \textit{bitrate} is important to control network congestion.
          \item[\textbf{energy\sul cons}] Edge devices are restricted in terms of resources and thus must economize or limit their energy \textit{consumption} while ensuring compliance with the remaining system requirements (i.e., other SLOs). 
          \item[\textbf{within\sul time}]Video processing introduces a considerable streaming \textit{delay}, which can lead to dropping frames and consequently poorer QoE. Hence, the stream's \textit{fps} can be adjusted to limit/avoid dropping frames.
          \item[\textbf{pixel\sul distance}] Measures the quality of the object tracking capacity; we expect the tracked object not to jump, but to have a smooth trajectory. Hence, we define a range for the acceptable \textit{distance}.
          \item[\textbf{transf\sul success}] Private or confidential information must not be disclosed; therefore, it must maximize the number of transformed faces in the stream.
    \end{description}


The workload was executed on the Jetson Xavier NX, which supports GPU-accelerated video processing over NVIDIA CUDA.
To explore correlations between \textit{GPU} and other metrics, we execute the entire workload twice on the Xavier NX -- once with and once without CUDA acceleration enabled.


\subsection{Model Construction}
\label{subsec:model-construction}

\begin{figure}[t]
\centering
\subfloat[energy\sul cons]{\includegraphics[width=.2\textwidth]{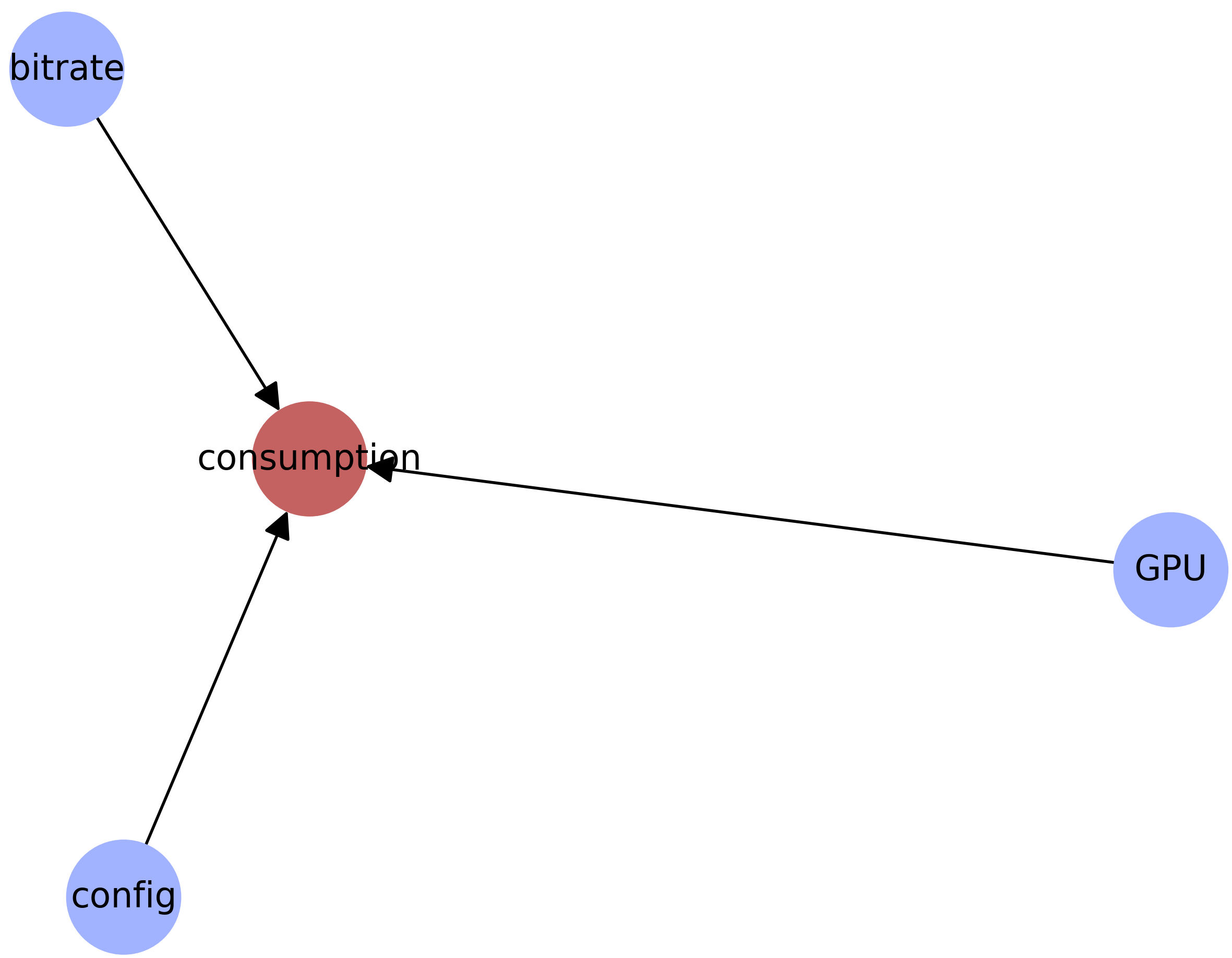}}
\subfloat[network\sul use]{\includegraphics[width=.2\textwidth]{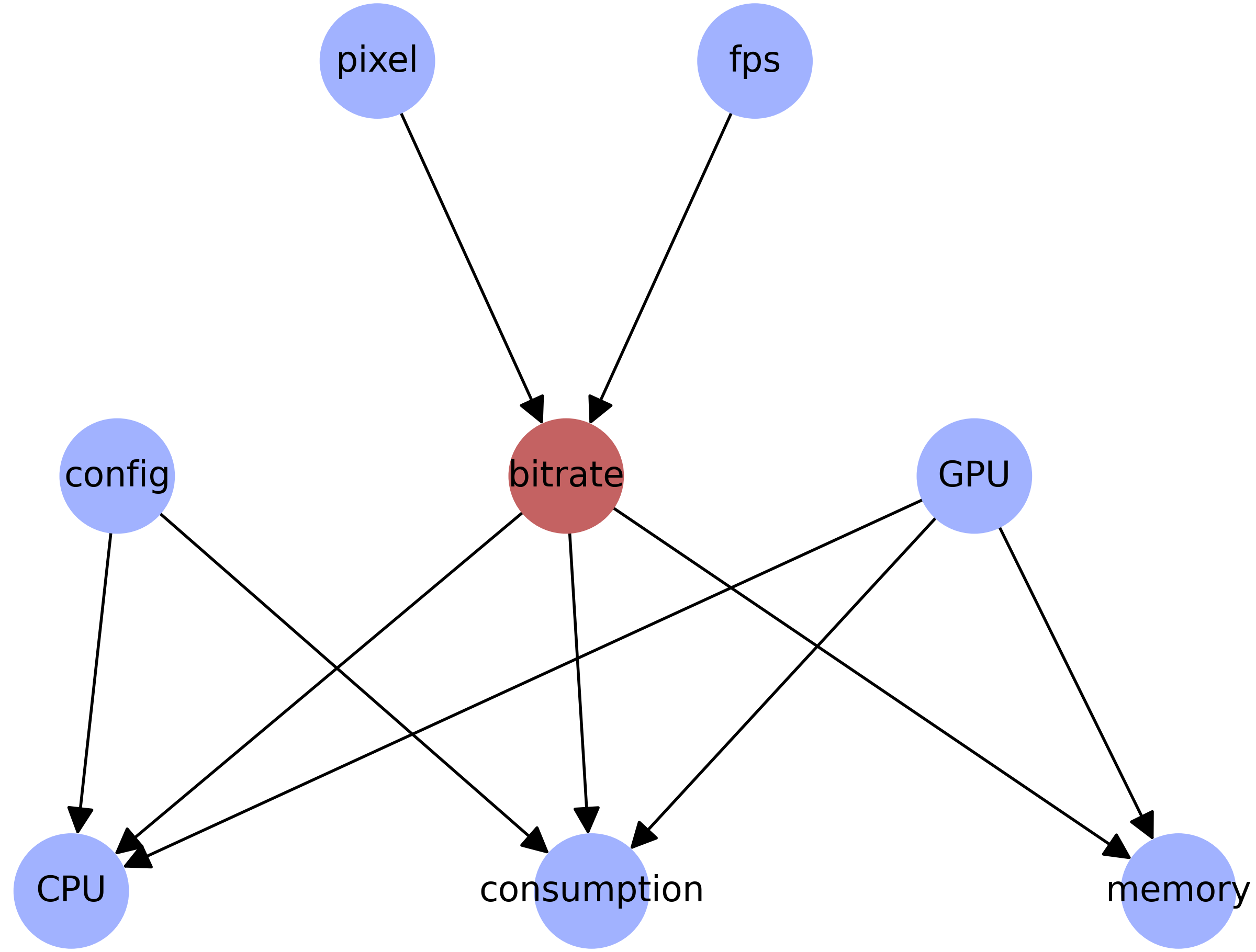}}
\subfloat[transf\sul succ]{\includegraphics[width=.2\textwidth]{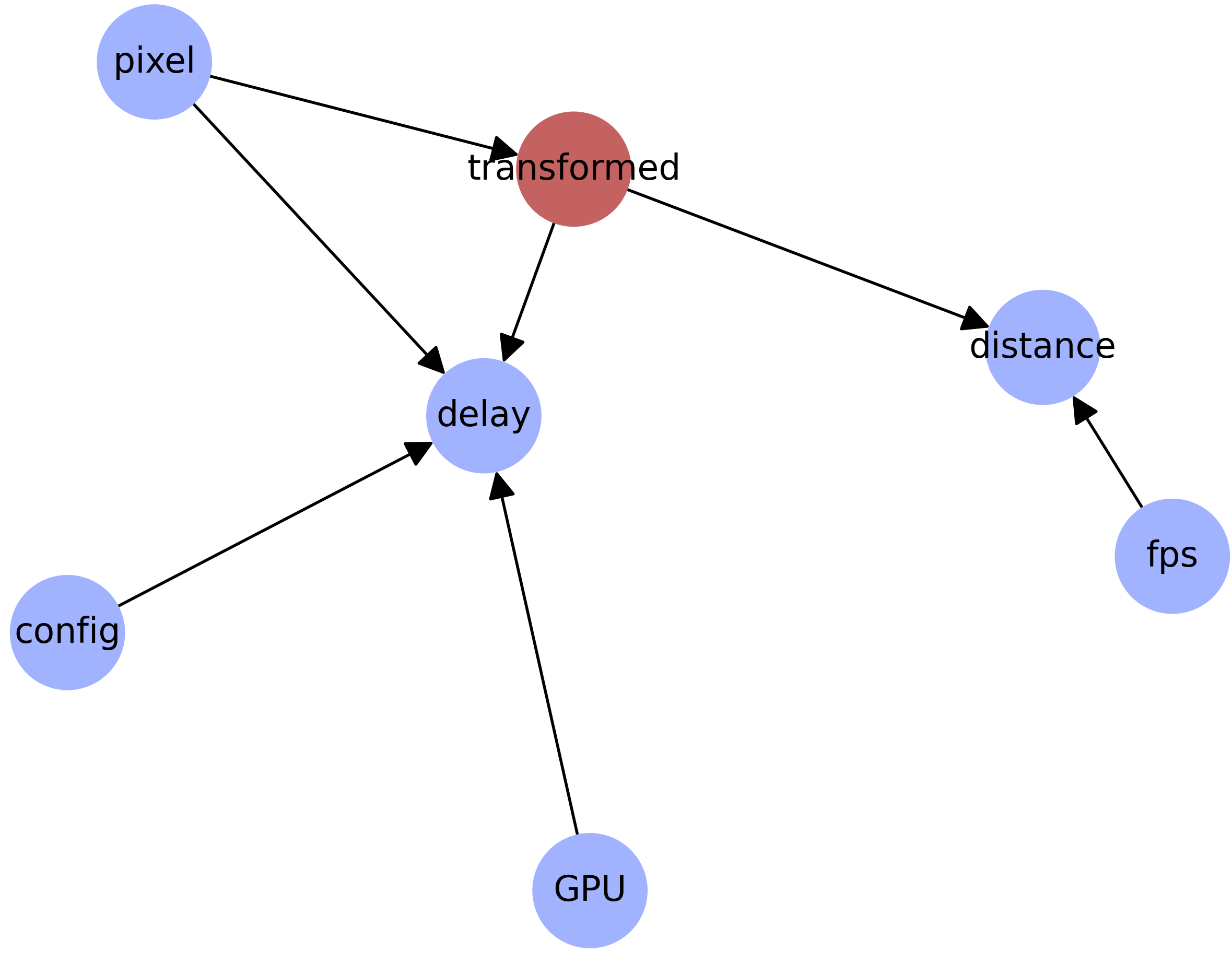}}
\subfloat[\textit{fps}]{\includegraphics[width=.2\textwidth]{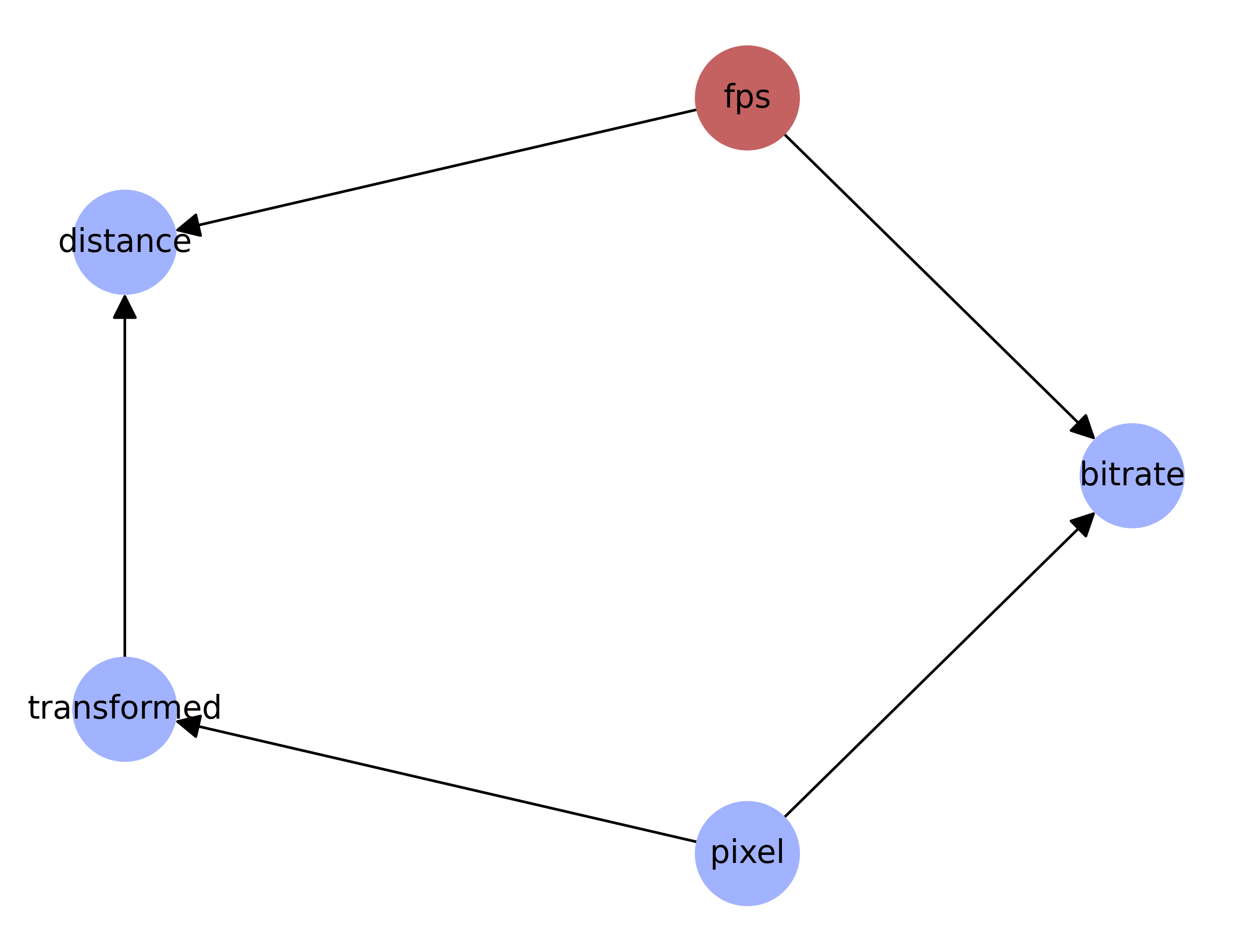}}
\subfloat[\textit{delay}]{\includegraphics[width=.2\textwidth]{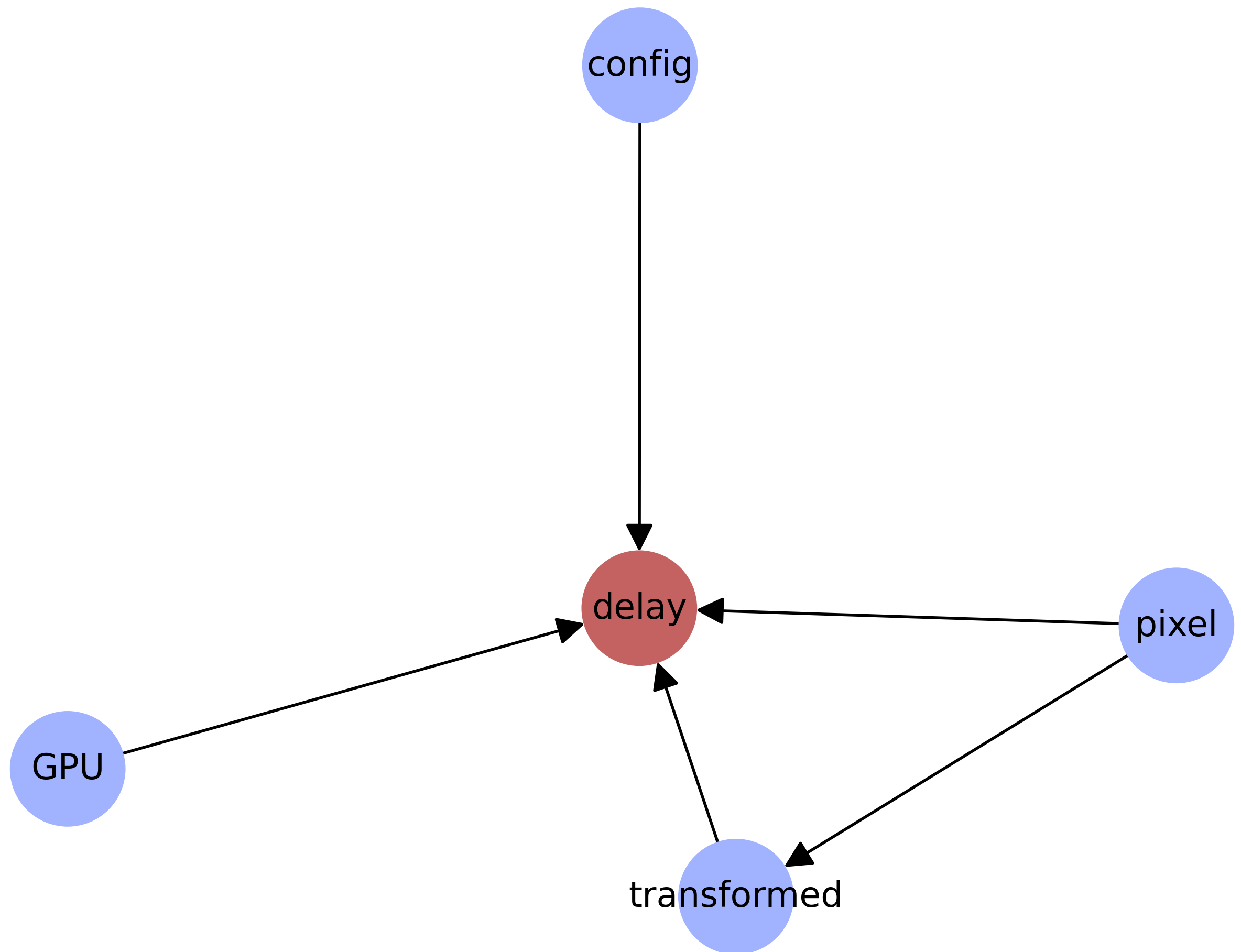}}
\caption{Markov blankets of the SLOs extracted from the Bayesian network}
\label{fig:slo-mbs}
\end{figure}

For constructing the BN, we leverage \textit{pgmpy}\footnote{https://pgmpy.org/, accessed June 14th, 2023}, a Python-based framework that supports an ample set of algorithms for structure and parameter learning, e.g., Hill-Climb Search (HCS) and Maximum Likelihood Estimation (MLE).
We train the BN with HCS and MLE on all captured metrics, which takes roughly 30 seconds on the Xavier NX.
%
%
%
After training the BN, we extract the MB for each SLO;
the resulting MBs are visualized in Figures~\ref{fig:slo-mbs}: The first three graphics show simple SLOs, i.e., such that require exactly one metric for evaluation. For example, \textbf{energy\sul cons} must evaluate \textit{consumption} to determine the state of the SLO. Metrics that have an edge pointing to \textit{consumption} (i.e., \textit{bitrate}, \textit{config}, and \textit{GPU}) causally influence the variable and thus the SLO fulfillment. On the other hand, \textbf{within\sul time}, is composed of two metrics and thus features two MBs.
Complex SLOs \cite{nastic_sloc_2020} (i.e., such that consist of $n$ metrics) would produce $n$ MBs; therefore, increasingly complex SLOs will require a sophisticated mechanism to merge and compress MBs.

We argue that the MBs extracted for each SLO are plausible because contained edges can be rationally explained.
Further, all MB SLOs contain at least one parameterizable metric within their sensory state, i.e., among the variables that influence the SLO outcome.
From a requirements perspective, this is essential because it allows a device to adapt dynamically to fulfill given SLOs.

\subsection{Device Configuration Inference}

To infer device configurations that comply with the SLOs, we extract information from the BN with Variable Elimination (VE). Instead of querying the entire BN, we execute the queries on the minimum subset of relevant variables, i.e., the MB of each SLO.
Since the MBs of the SLOs contained all three parameterizable metrics (i.e., \textit{fps}, \textit{pixel}, and \textit{config}), a device must include these parameters in an inferred configuration; otherwise, there is no full control over the SLOs. However, suppose we would only trace a subset of the SLOs (e.g., \textbf{network\sul usage} \& \textbf{transf\sul success}), a configuration must only include the respective parameters contained in the MBs, e.g., \textit{fps} \& \textit{pixel}, but not \textit{config}.

VE computes the probability of SLO violations for exactly one parameter assignment; we repeatedly apply this approach for all assignments. To be precise, the parameter space for (\textit{pixel} : \textit{fps} : \textit{config}) consists of (5 : 6 : 3) possible assignments.
%
Iterating over $5*6*3=90$ combinations and $5$ SLO-MBs produces $5 * 90 = 450$ inference queries, which require roughly 700ms on the Jetson Xavier NX.
The result is a list of configurations that fulfill the given SLOs, e.g., one could be (240p : 20fps : 4C\_20W).
To deal with changing requirements and heterogeneous characteristics of CC devices, it is possible to provide additional constraints to the VE (e.g., \textit{GPU}=False), or customize SLOs to rank a metric rather than limiting it (e.g. minimize \textit{consumption}).

\subsection{Evaluation}

To evaluate the quality of inferred configurations, we compare the number of SLO violations between devices that apply inferred or arbitrary configurations.
%
We envision two scenarios that are based on the workload for face blurring. The scenarios are described below, while the corresponding SLO thresholds are presented in Table~\ref{tab:slo-thresholds}. We intend to minimize \textbf{energy\sul cons} for both scenarios regardless of whether the energy supply would be constrained:
\vspace{4pt}

\noindent\textbf{Scenario A:} To create a virtual map (like  Google Street View\footnote{https://www.google.com/streetview/, accessed June 18th 2023}), a camera-equipped car captures street videos. The car has an edge device installed to transform the stream; the result is directly rendered to a local map and only accessed remotely in case of inspection, so \textbf{network\sul usage} is of less importance. We assume the rendering process to run in the background; therefore, the GPU is not available for processing.
To create a detailed map, \textbf{pixel\sul distance} must be low, and \textbf{within\sul time} fulfilled in most cases. However, the stream can be re-rendered to blur undetected faces, thus \textbf{transf\sul success} is less critical.

\noindent\textbf{Scenario B:} Within a smart factory, employees equipped with head-mounted cameras conduct an audit. To protect privacy, the video stream is transformed on an edge device before streaming to remote consumers. Video content is intended for live inspection only; therefore, \textbf{pixel\sul distance} and \textbf{within\sul time} are less important, while high \textbf{transf\sul success} prevents privacy breaches. However, since audits involve various providers and consumers, low \textbf{network\sul usage} is desired.


\setlength{\tabcolsep}{3pt}
\begin{table}[t]
  \centering
  \caption{SLO thresholds that reflect the scenarios' requirements}
  \label{tab:slo-thresholds}
  \resizebox{\columnwidth}{!}{
  \begin{tabular}{ccccccc}
    \toprule
    Scenario  & \textbf{transf\sul success} & \textbf{distance} &  \textbf{network\sul usage}  &  \textbf{within\sul time} & \textbf{energy\sul cons} & GPU\\
    \midrule
    A & $\geq 90\%$ & $\leq 35$ & $\leq 8.2\;Mio.\;px/s$ & $\geq 95\%$ & $min(x)$ & No \\
    \midrule
    B & $\geq 98\%$ & $\leq 60$ & $\leq 1.6\;Mio.\;px/s$ & $\geq 75\%$ & $min(x)$ & Yes\\
    \bottomrule
  \end{tabular}
  }
\end{table}

\setlength{\tabcolsep}{5pt}
\begin{table}[t]
  \centering
  \caption{List of configurations generated by exact inference or picked naively}
  \label{tab:configuration-list}
  \begin{tabular}{cccccc}
    \toprule
    Scenario & Source & Resolution & FPS &  Mode  &  GPU\\
    \midrule
    \multirow{4}{5em}{\hspace{15pt} A}   & inferred   & 240p & 20    & 4C\_15W & \multirow{4}{5em}{\hspace{14pt} No} \\
                                         & naive      & 360p & 30    & 6C\_20W & \\
                                         & random \#1 & 120p & 16    & 6C\_20W & \\
                                         & random \#2 & 720p & 12    & 2C\_10W & \\
    \midrule
    \multirow{4}{5em}{\hspace{15pt} B}   & inferred   & 240p & 16    & 2C\_10W & \multirow{4}{5em}{\hspace{13pt} Yes} \\
                                         & naive      & 180p & 26    & 4C\_15W & \\
                                         & random \#1 & 360p & 20    & 2C\_15W & \\
                                         & random \#2 & 480p & 30    & 6C\_20W & \\
    \bottomrule
  \end{tabular}
\end{table}

We supply the SLO thresholds to the inference mechanism; the resulting configurations are presented in Table~\ref{tab:configuration-list}: The first line contains the inferred configuration, and the second line the naive assumption; the third and fourth lines are randomly generated.
To evaluate the number of SLO violations, we measured each configuration's performance over 10 min; results are presented in Table~\ref{tab:results}. Over the measurement course, the inferred configurations fulfilled the SLOs for both scenarios. The naive assumption, on the other hand, violated one SLO within each scenario (red cell), i.e., in Scenario A it failed to fulfill \textbf{within\sul time}, while in Scenario B \textbf{transf\sul success} was violated. The randomly generated configurations committed two SLO violations in Scenario A and one in Scenario B each.
%
The results show that our inferred configurations fulfilled all given SLOs while also consuming the least energy.

\setlength{\tabcolsep}{5pt}
\begin{table}[t]
  \centering
  \caption{Fulfillment of SLOs depending on scenario and configuration}
  \label{tab:results}
  \resizebox{\columnwidth}{!}{
  \begin{tabular}{ccccccc}
    \toprule
    Scenario & Source & \textbf{transf\sul success} & \textbf{distance} &  \textbf{network\sul usage}  &  \textbf{within\sul time} & \textbf{energy\sul cons}\\
    \midrule
    \multirow{4}{5em}{\hspace{15pt} A}   & inferred   & 98\%  & 15 (97\%) & 2.0 Mio. & 100\% & \cellcolor{green!20}6.0W\\
                                         & naive      & 100\% & 10 (100\%)& 6.9 Mio. & \cellcolor{red!25}92\% & 8.0W\\
                                         & random \#1 & \cellcolor{red!25}4\%   & \cellcolor{red!25}127 (2\%) & 0.4 Mio. & 100\% & 7.0W\\
                                         & random \#2 & 100\% & \cellcolor{red!25}28 (89\%) & \cellcolor{red!25}11 Mio. & 100\% & \cellcolor{green!20}6.0W\\
    \midrule
    \multirow{4}{5em}{\hspace{15pt} B}   & inferred   & 98\%  & 18(98\%) & 1.6 Mio. & 100\% & \cellcolor{green!20}6.0W\\
                                         & naive      & \cellcolor{red!25}92\% & 11(99 \%) & 1.5 Mio. & 100\% & 6.5W\\
                                         & random \#1 & 99\% & 15 (100\%)& \cellcolor{red!25}4.6 Mio. & 100\% & \cellcolor{green!20}6.0W\\
                                         & random \#2 & 100\% & 10 (100\%) & \cellcolor{red!25}12.3 Mio. & 97\% & 7.5W\\
                                         
    \bottomrule
  \end{tabular}
  }
\end{table}



\section{Conclusion \& Future Work}
\label{sec:conclusion}
This paper proposed a statistical reasoning model for explaining causal relations between metrics and the system state, which is reflected by a set of SLOs and their MBs.
%
%
Essentially, this provides individual edge devices with decentralized intelligence, which helps to cope with the scale and complexity of CC systems.
Our methodology was able to provide configurations that would not commit SLO violations; however, the scale of CC systems makes it necessary to assess the impacts of increasingly large Bayesian networks in terms of performance and precision. Furthermore, to cover heterogeneity among CC devices, we aim to infer configurations for arbitrary devices.

\bibliographystyle{splncs04}
\bibliography{boris,praveen,victor}

\begin{thebibliography}{10}
\providecommand{\url}[1]{\texttt{#1}}
\providecommand{\urlprefix}{URL }
\providecommand{\doi}[1]{https://doi.org/#1}

\bibitem{aliferis_short_2010}
Aliferis, C., et~al.: Local {Causal} and {Markov} {Blanket} {Induction} for
  {Causal} {Discovery} and {Feature} {Selection} for {Classification} {Part}
  {I}: {Algorithms} and {Empirical} {Evaluation}. Journal of Machine Learning
  Research  \textbf{11},  171--234 (Jan 2010)

\bibitem{baniya_short_2020}
Baniya, P., et~al.: Towards {Policy}-aware {Edge} {Computing} {Architectures}.
  In: 2020 {IEEE} {International} {Conference} on {Big} {Data} ({Big} {Data})
  (Dec 2020)

\bibitem{Beckman2020}
Beckman, P., et~al.: Harnessing the computing continuum for programming our
  world. In: Fog {Computing}, pp. 215--230. John Wiley \& Sons, Ltd (Apr 2020)

\bibitem{bui_learning_2012}
Bui, A., Jun, C.H.: Learning {Bayesian} {Network} {Structure} {Using} {Markov}
  {Blanket} {Decomposition}. Pattern Recognition Letters  \textbf{33} (Dec
  2012)

\bibitem{casamayor_pujol_towards_2021}
Casamayor~Pujol, V., Raith, P., Dustdar, S.: Towards a new paradigm for
  managing computing continuum applications. In: {IEEE} 3rd {International}
  {Conference} on {Cognitive} {Machine} {Intelligence}, {CogMI} 2021. pp.
  180--188 (2021)

\bibitem{cheng_short_2002}
Cheng, J., et~al.: Learning {Bayesian} networks from data: {An}
  information-theory based approach. Artificial Intelligence
  \textbf{137}(1-2),  43--90 (May 2002)

\bibitem{donta_governance_2023}
Donta, P.K., Sedlak, B., Casamayor~Pujol, V., Dustdar, S.: Governance and
  sustainability of distributed continuum systems: a big data approach. Journal
  of Big Data  \textbf{10}(1), ~53 (Apr 2023). \doi{10.1186/s40537-023-00737-0}

\bibitem{dustdar_principles_2011}
Dustdar, S., Guo, Y., Satzger, B., Truong, H.L.: Principles of {Elastic}
  {Processes}. Internet Computing, IEEE  \textbf{15},  66--71 (Nov 2011)

\bibitem{dustdar_distributed_2023}
Dustdar, S., Pujol, V.C., Donta, P.K.: On {Distributed} {Computing} {Continuum}
  {Systems}. IEEE Transactions on Knowledge and Data Engineering
  \textbf{35}(4),  4092--4105 (Apr 2023). \doi{10.1109/TKDE.2022.3142856}

\bibitem{friston_life_2013}
Friston, K.: Life as we know it. Journal of The Royal Society Interface
  \textbf{10}(86),  20130475 (Sep 2013). \doi{10.1098/rsif.2013.0475}

\bibitem{furst_elastic_2018}
Fürst, J., Fadel~Argerich, M., Cheng, B., Papageorgiou, A.: Elastic {Services}
  for {Edge} {Computing}. In: 2018 14th {International} {Conference} on
  {Network} and {Service} {Management} ({CNSM}). pp. 358--362 (Nov 2018), iSSN:
  2165-963X

\bibitem{kirchhoff_markov_2018}
Kirchhoff, M., Parr, T., Palacios, E., Friston, K., Kiverstein, J.: The
  {Markov} blankets of life: autonomy, active inference and the free energy
  principle. Journal of The Royal Society Interface  \textbf{15}(138),
  20170792 (Jan 2018)

\bibitem{nastic_sloc_2020}
Nastic, S., Morichetta, A., Pusztai, T., Dustdar, S., Ding, X., Vij, D., Xiong,
  Y.: {SLOC}: {Service} {Level} {Objectives} for {Next} {Generation} {Cloud}
  {Computing}. IEEE Internet Computing  \textbf{24}(3) (May 2020).
  \doi{10.1109/MIC.2020.2987739}

\bibitem{palacios_markov_2020}
Palacios, E.R., Razi, A., Parr, T., Kirchhoff, M., Friston, K.: On {Markov}
  blankets and hierarchical self-organisation. Journal of Theoretical Biology
  \textbf{486} (Feb 2020)

\bibitem{pearl_probabilistic_1988}
Pearl, J.: Probabilistic reasoning in intelligent systems : networks of
  plausible inference. San Mateo, Calif. : Morgan Kaufmann Publishers (1988)

\bibitem{prentice_smart_2018}
Prentice, C.T., Karakonstantis, G.: Smart {Office} {System} with {Face}
  {Detection} at the {Edge}. In: 2018 {IEEE} {SmartWorld}. pp. 88--93 (Oct
  2018)

\bibitem{scanagatta2019survey}
Scanagatta, M., Salmer{\'o}n, A., Stella, F.: A survey on bayesian network
  structure learning from data. Progress in Artificial Intelligence
  \textbf{8},  425--439 (2019)

\bibitem{sedlak_controlling_2023}
Sedlak, B., Casamayor~Pujol, V., Donta, P.K., Dustdar, S.: Controlling {Data}
  {Gravity} and {Data} {Friction}: {From} {Metrics} to {Multidimensional}
  {Elasticity} {Strategies}. In: {IEEE} {SSE} 2023. Chicago, IL, USA (Jul
  2023), (Accepted)

\bibitem{sedlak_privacy_2023}
Sedlak, B., Murturi, I., Donta, P.K., Dustdar, S.: A {Privacy} {Enforcing}
  {Framework} for {Transforming} {Data} {Streams} on the {Edge}. IEEE
  Transactions on Emerging Topics in Computing  (2023), (Under submission)

\bibitem{tsamardinos_time_2003}
Tsamardinos, I., Aliferis, C.F., Statnikov, A.: Time and sample efficient
  discovery of {Markov} blankets and direct causal relations. Association for
  Computing Machinery, New York, USA (Aug 2003). \doi{10.1145/956750.956838}

\bibitem{tarneberg_6g_2022}
Tärneberg, W., et~al.: The {6G} {Computing} {Continuum} ({6GCC}): {Meeting}
  the {6G} computing challenges. In: {International} {Conference} on {6G}
  {Networking} (Jul 2022)

\bibitem{wang_towards_2020}
Wang, H., Ling, Z., Yu, K., Wu, X.: Towards efficient and effective discovery
  of {Markov} blankets for feature selection. Information Sciences
  \textbf{509} (Jan 2020)

\bibitem{zhang_simple_1994}
Zhang, N., Poole, D.: A simple approach to {Bayesian} network computations. In:
  Engineering-{Economic} {Systems}, {Stanford} {University} (1994)

\end{thebibliography}
\end{document}